\documentclass[intlimits,twoside,a4paper]{article}
\usepackage[cp1251]{inputenc}

\usepackage[eqsecnum]{cmpj3}

\usepackage{bm}
\usepackage{graphicx}

\issue{2019}{22}{4}{43601}
\doinumber{10.5488/CMP.22.43601}
\title{Neutron scattering investigations of the boson peak}%
\author[U. Buchenau]{U. Buchenau}
\address{Forschungszentrum J\"ulich GmbH, J\"ulich Centre for Neutron Science (JCNS-1) and Institute for Complex Systems (ICS-1), 52425 J\"ulich, Germany}

\date{Received June 6, 2019, in final form July 31, 2019}

\begin{document}

\maketitle

\begin{abstract}
Inelastic neutron scattering is not only capable of determining a generalized vibrational density of states around the boson peak of a glass, but can also be used to get information on the eigenvectors. The eigenvectors determine the dynamic structure factor, which is $Q^2S(Q)$ ($Q$ wavevector) for long wavelength sound waves. This enables the determination of the sound wave fraction below and at the boson peak, done for SiO$_2$, B$_2$O$_3$, polybutadiene and amorphous germanium. The temperature dependence of the boson peak in silica and glycerol is shown. X-ray Brillouin scattering data show that the damping of the longitudinal sound waves in silica and glycerol follows the soft potential model $\omega^4$ prediction in the limit of low frequency.  
\pacs 61.20.Lc, 61.43.Fs, 61.12.Ex

\end{abstract}

\section{Introduction}

An important part of Giancarlo Ruocco's scientific work was devoted to the riddle of the low frequency vibrations in glasses. Glasses have not only the low frequency longitudinal and transverse sound waves which one expects in an elastic solid on the basis of the Debye model, but also additional excitations, tunneling states and soft vibrations, which are clearly visible in the heat capacity, the thermal conductivity and the sound absorption at low temperatures \cite{philbook}.

In scattering techniques (Raman, neutron and x-ray scattering), the additional soft vibrations appear as a broad peak between the frequency zero and the lowest Van Hove singularity of the corresponding crystal. At lower temperatures, it shows the linear temperature increase expected for harmonic vibrations, so it got the name boson peak. In the heat capacity ($c_\text{p}$) measurements, it appears as a peak in $c_\text{p}/T^3$ around 5 to 10 K. Thus, the boson peak is not a peak in the density of states $g(\omega)$, but rather a peak in $g(\omega)/\omega^2$, which is a constant in the Debye model.   

The present paper summarizes the neutron scattering studies of the boson peak. As will be seen, neutrons are a good tool to study the vibrational density of states of a glass and are capable of providing information on the mode eigenvectors. However, if it comes to Brillouin scattering studies of the longitudinal sound absorption in the THz range, x-rays are much better than neutrons, because their energy loss or gain in the Brillouin scattering process is negligible. This advantage was established by the first x-ray Brillouin measurements in vitreous silica \cite{benassi1996,foret1996}, for Giancarlo Ruocco the start of an impressive scientific career.

The paper focuses on the neutron results in section~\ref{sec2} and the low frequency limit of neutron and x-ray data in section~\ref{sec3}. Section~\ref{sec4} contains the conclusions.

\section{Neutron scattering at the boson peak}\label{sec2}

The theoretical basis for  neutron scattering studies of the boson peak was supplied by Carpenter and Pelizzari \cite{carp}, who treated the scattering from low frequency sound waves in an elastically isotropic solid. They separated the Brillouin scattering from longitudinal sound waves at low wavevector $Q$ (that is, the one where x-rays are much better suited than neutrons) from the $Q^2S(Q)$-signal at higher wavevectors, the Umklapp scattering, which contains contributions from all sound waves.

A nice feature of the Umklapp scattering from sound waves is that its intensity at a known temperature is given by the Debye density of states and $S(Q)$, more precisely the elastic scattering $S(Q)\exp(-2W)$, which has the same Debye-Waller factor as the one-phonon scattering.

\begin{figure}[!b]
\centering
\includegraphics[width=.54\textwidth]{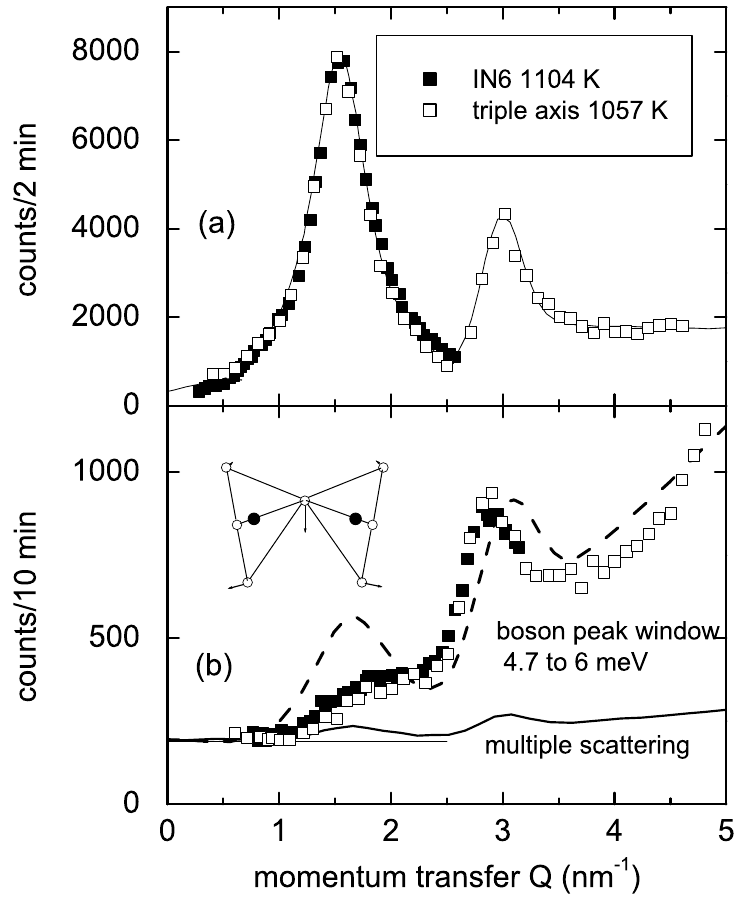}
\caption{Neutron scattering from vitreous silica at higher temperatures as a function of the wavevector~$Q$~\cite{wischi1998} (a) Elastic scattering (b) Inelastic scattering in the boson peak window. There is a $Q$-independent multiple scattering background, which can be disregarded. The sound wave scattering is low (the continuous line). When multiplied with 6, one sees that it still does not describe the data, which have the dynamic structure factor of coupled rigid SiO$_4$-tetrahedra.} 
\label{fig1}
\end{figure}

If the low frequency vibrations are not sound waves, their scattering and its wavevector dependence can be calculated from their density of states and their eigenvector. This property has been used to identify the modes at the boson peak of vitreous silica as coupled librations of SiO$_4$-tetrahedra (see figure~\ref{fig1}).  

Vitreous silica happens to have a very impressive boson peak, with an intensity which is several times higher than the Debye level. Figure~\ref{fig1}~(b) compares the measured data \cite{wischi1998} in the boson peak frequency window to the predictions of Carpenter and Pelizzari \cite{carp} (the continuous line) for the sound wave scattering. It is clearly seen that the measured signal is not only a factor of four to five higher than the sound wave prediction, but that it also has a strongly different wavevector dependence (the dashed line shows the sound wave expectation multiplied by the factor six). It turns out that the measured wavevector dependence is well described by coupled librations of rigid SiO$_4$-tetrahedra, a finding which has been corroborated in later experimental \cite{hehlen} and numerical \cite{chumakov} work.

The dynamic structure factor of the boson peak vibrations has been also measured in boron trioxide~\cite{engberg} and in polybutadiene \cite{prl1996}. In these two cases, one finds a mixture of the sound wave structure factor with a simple $Q^2$-dependence, with the sound wave structure factor fraction extrapolating to the Debye density of states at zero frequency.

An interesting case is amorphous germanium, which does not have a boson peak and markedly less tunneling states than other glasses \cite{neufville}. Instead of a boson peak, it has a low frequency tail of the lowest Van Hove singularity of the crystal at 2 THz. In this tail, neutron data  \cite{prager} find a mixture of sound waves and of the bond bending motion of the lowest Van Hove singularity.

\begin{figure}[!t]
\centering
\includegraphics[width=.55\textwidth]{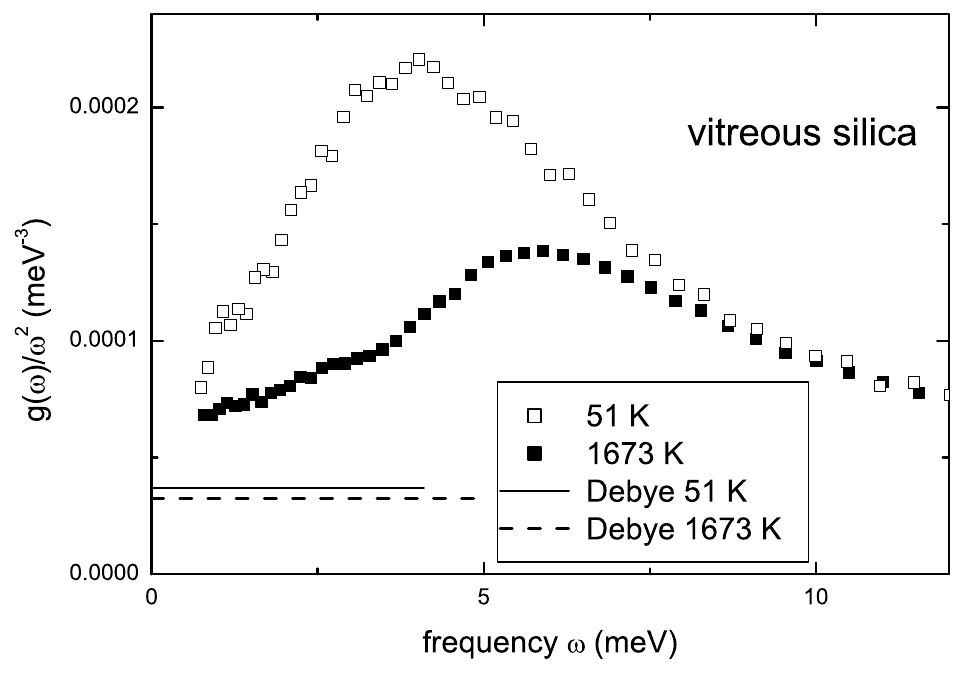}
\caption{The boson peak in silica at 51 and 1673 K \cite{wischi1998} (the glass transition is at 1473 K).} 
\label{fig2}
\end{figure}

The boson peak in vitreous silica has an unusual temperature dependence (see figure~\ref{fig2}), with a strong hardening at increasing temperature \cite{wischi1998}, being continuous even through the glass transition at 1473~K. Figure~\ref{fig2} compares the measured vibrational density of states \cite{wischi1998} at 51 and 1673~K.

\begin{figure}[!b]
\centering
\includegraphics[width=.55\textwidth]{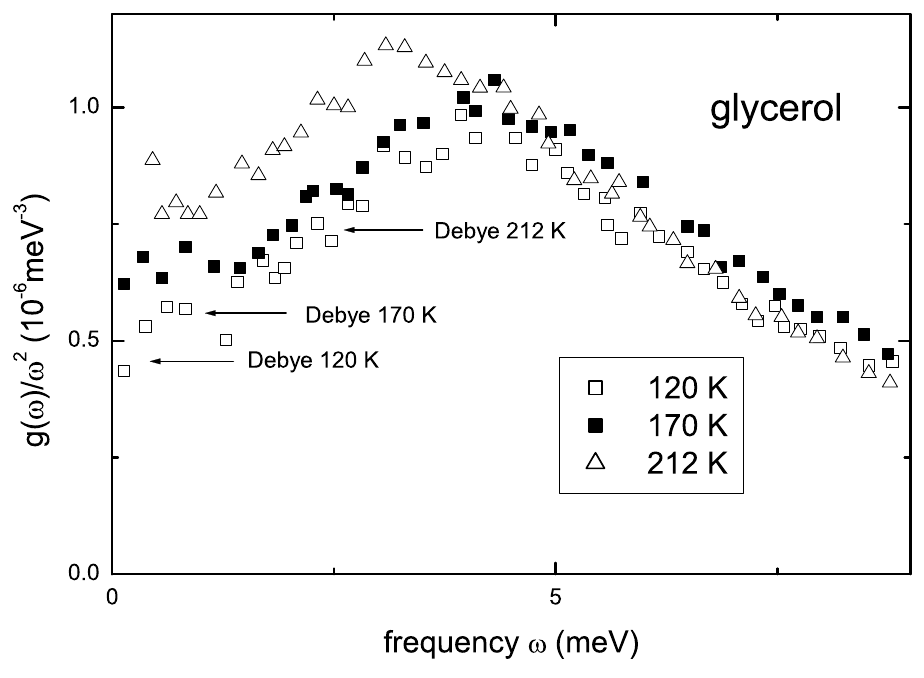}
\caption{The boson peak in glycerol at 120, 170 and 212 K \cite{wuttke} (the glass transition is at 187 K).} 
\label{fig3}
\end{figure}

The usual temperature dependence of the boson peak, namely a softening with increasing temperature which becomes even more marked above the glass transition temperature $T_\text{g}$, is demonstrated in figure~\ref{fig3} for glycerol \cite{wuttke}. 

\section{Low frequency limit of neutron and x-ray Brillouin data} \label{sec3}

Though both neutron and x-ray inelastic scattering data are limited by the instrumental resolution at low frequency, it has been possible to obtain information in substances with a relatively high boson peak frequency. Figure~\ref{fig2} and figure~\ref{fig3} show that for low enough temperature the density of states extrapolates to the Debye density of states, i.e., that the density $g_\text{s}(\omega)$ of the additional soft modes tends to zero. 

The way in which $g_\text{s}(\omega)$ disappears is better seen in heat capacity data \cite{zeller,talon}, which show a $T^5$-increase, compatible with $g_\text{s}(\omega)\propto\omega^4$.

This conclusion has been recently corroborated by numerical results \cite{lerner1,lerner2,wang}. In the first of these \cite{lerner1}, it was observed that all these localized soft modes have a positive fourth order term in their displacement potential. Both findings, a $g_\text{s}(\omega)\propto\omega^4$ and a positive fourth order term in the mode potential, are predictions of the soft potential model \cite{ramos} for the low temperature anomalies of glasses, a model which considers tunneling states and soft vibrations as members of a common distribution.

For the x-ray Brillouin data, the soft potential model predicts a damping $\Gamma\propto\omega^4$, which results from the rise of $g_\text{s}(\omega)$
\begin{equation}\label{gamom}
\Gamma(\omega)={\rm v}_ll^{-1}_\text{res,vib}=\frac{\piup\omega C_l}{8}\left(\frac{\hbar\omega}{W}\right)^3=\frac{\omega^4}{2\piup\omega_\text{IR}^3}\,,
\end{equation}
where $\Gamma(\omega)$ is the half width at half maximum of the Brillouin line, ${\rm v}_l$ is the longitudinal sound velocity, $l^{-1}_\text{res,vib}$ is the mean free path of the longitudinal waves under the resonant interaction with the soft vibrational modes, $C_l$ is a dimensionless constant, and $W$ is the crossover energy between tunneling states and soft vibrations. Equation~(\ref{gamom}) defines the Ioffe-Regel frequency $\omega_\text{IR}$, where $\Gamma=\omega/2\piup$ and the mean free path is equal to the wavelength. Table~\ref{tab1} shows that the soft potential fits of the plateau in the thermal conductivity for silica \cite{ramos} and for glycerol \cite{talon} predict the Ioffe-Regel limit at the boson peak for both glasses.

\begin{table}[!t]
	\centering
\caption{Soft potential parameters and Ioffe-Regel limit for vitreous silica \cite{ramos} and glycerol \cite{talon}.}
		\label{tab1}
\vspace{2ex}
		\begin{tabular}{|c|c|c|c|c|}
\hline\hline 
substance   & $W/k_\text{B}$, K      &   $C$, $ \times10^{4}$      &   $v_l$, km/s   &  $\hbar\omega_\text{IR}$, meV   \\
\hline   
\hline
silica      & 3.9          &   2.6      &   5.8    &     3.9              \\
glycerol    & 4.3          &   1.9      &   3.6    &     4.8              \\
\hline\hline 
		\end{tabular}
\end{table}

\begin{figure}[!b]
\centering
\includegraphics[width=.55\textwidth]{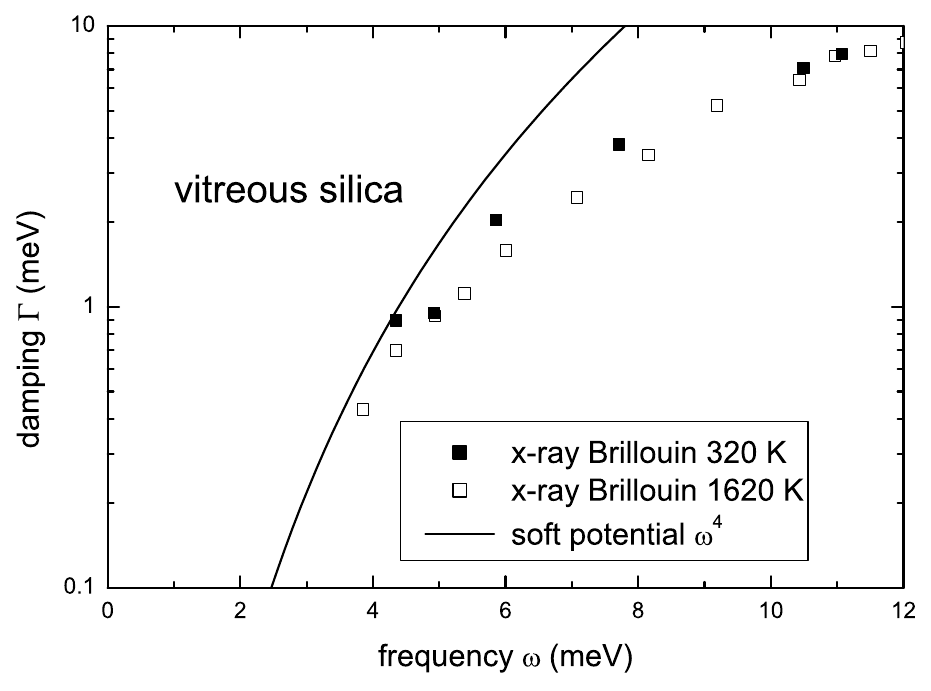}
\caption{Comparison of measured x-ray Brillouin damping \cite{baldi} in vitreous silica at 320 and 1620 K with the soft potential expectation of equation~(\ref{gamom}).} 
\label{fig4}
\end{figure}

An alternative explanation of a damping $\Gamma\propto\omega^4$ is Rayleigh scattering from the static disorder. However, this explanation was found to be too weak to explain the strong sound wave damping in glasses~\cite{kittel}. 

What one usually observes in x-ray Brillouin data, however, is not an $\omega^4$-increase (or $Q^4$-increase, where $Q$ is the Brillouin wavevector) of $\Gamma$, but rather $\Gamma\propto Q^2$, a behavior which was explained by Walter Schirmacher, Giancarlo Ruocco and Tullio Scopigno \cite{srs} in terms of a gaussian distribution of local shear restoring forces around the average one, based on Schirmachers theory \cite{s} of the plateau in the thermal conductivity.

However, if one goes low enough in frequency, one begins to see deviations from the $Q^2$-dependence. The first indications for an $\omega^4$-dependence of $\Gamma$ were found in densified silica \cite{ruffle1} and lithium borate~\cite{ruffle2}, where one does not have soft potential fits. However, recent x-ray Brillouin scattering data \cite{baldi} managed to find clear indications for a $Q^4$-dependence of the damping $\Gamma$ of the longitudinal sound waves in silica at the lowest accessible frequencies at 300 and 1620 K. These data are compared with the soft potential prediction of equation~(\ref{gamom}) in figure~\ref{fig4}, taking the soft potential parameters from table~\ref{tab1}.

The soft potential paper \cite{ramos} does only determine an average coupling constant for longitudinal and transverse sound waves, but for vitreous silica one knows from measurements \cite{berret} below 1 K that $C$ is the same for longitudinal and transverse waves.
 
In the case of glycerol, the x-ray Brillouin data \cite{pnas} show a clearly defined $Q^4$-damping, which extrapolates to the Ioffe-Regel limit at 5.2 meV, in close agreement with the soft potential value of 4.8~meV in table~\ref{tab1}.

\section{Conclusions} \label{sec4}

In the foregoing, it was demonstrated that x-ray and neutron scattering can provide ample and complementary information on the vibrational soft modes of the boson peak. While x-rays are better capable of measuring the width and the position of the Brillouin peaks, neutrons suit  well to measure the vibrational density of states as well as its temperature dependence. In addition, they are capable of providing information on the eigenvectors, as demonstrated in section~\ref{sec2} for vitreous silica.

The ongoing improvement of the x-ray Brillouin technique has made it possible to measure the deviations from a damping proportional to $\omega^2$ at lower frequencies in vitreous silica \cite{baldi}. In the present paper, it was shown that these deviations extrapolate to the low-frequency prediction of the soft potential model. In glycerol, the measured $Q^4$-dependence of the x-ray Brillouin damping is close to the soft potential prediction. The agreement supports the hypothesis that the $Q^4$-damping in glasses is connected to the $\omega^4$-rise of the density of vibrational non-phonon modes, seen in the low-temperature heat capacity and in recent numerical work, related to the tunneling states as postulated in the soft potential model.

\ukrainianpart

\title{Дослідження бозонного піку методом розсіяння нейтронів}
\author{У. Бухенау}
\address{Дослідницький центр Юліх, Юліхський центр нейтронних досліджень (JCNS-1) та Інститут складних систем (ICS-1), 52425 Юліх, Німеччина}

\makeukrtitle

\begin{abstract}

Непружне розсіювання нейтронів не лише дозволяє визначити узагальнену густину коливних станів біля бозонного піку
в скловидному стані, але також може бути використане для отримання інформації про власні вектори. Власні вектори
визначають динамічний структурний фактор, який є $Q^2S(Q)$ ($Q$ хвильовий вектор) для довгохвильових звукових хвиль.
Це дозволяє встановлення частки звукових хвиль нижче та на бозонному піку, що було зроблено для SiO$_2$, B$_2$O$_3$,
полібутадієну та аморфного германію. Показано температурну залежність бозонного піку в кремнеземі та гліцеролі.
Дані брілюенівського розсіювання рентгенівських променів вказують, що загасання повздовжних хвиль в кремнеземі та 
гліцеролі слідує передбаченням моделі м'яких потенціалів  $\omega^4$ в границі низької частоти.

\keywords  скловидні системи, непружне розсіювання нейтронів, бозонний пік 

\end{abstract}

\end{document}